\documentclass[sigconf]{acmart}
\AtBeginDocument{%
  }
\usepackage{listings}
\usepackage{multicol}

\copyrightyear{2024}
\acmYear{2024}
\setcopyright{acmlicensed}\acmConference[IUI '24]{29th International Conference on Intelligent User Interfaces}{March 18--21, 2024}{Greenville, SC, USA}
\acmBooktitle{29th International Conference on Intelligent User Interfaces (IUI '24), March 18--21, 2024, Greenville, SC, USA}
\acmDOI{10.1145/3640543.3645143}
\acmISBN{979-8-4007-0508-3/24/03}

\begin{document}

\title{LAVE: LLM-Powered Agent Assistance and Language Augmentation for Video Editing}








\author{Bryan Wang}
\authornote{The work was done during the author's internship at Meta Reality Labs - Research.}
\affiliation{%
  \institution{University of Toronto}
  \city{Toronto}
  \state{ON}
  \country{Canada}}
\email{bryanw@dgp.toronto.edu}

\author{Yuliang Li}
\affiliation{%
  \institution{Reality Labs Research, Meta}
  \city{Sunnyvale}
  \state{CA}
  \country{USA}}
\email{yuliangli@meta.com}

\author{Zhaoyang Lv}
\affiliation{%
  \institution{Reality Labs Research, Meta}
  \city{Sunnyvale}
  \state{CA}
  \country{USA}}
\email{zhaoyang@meta.com}

\author{Haijun Xia}
\affiliation{%
  \institution{University of California San Diego }
  \city{La Jolla}
  \state{CA}
  \country{USA}}
\email{haijunxia@ucsd.edu}

\author{Yan Xu}
\affiliation{%
  \institution{Reality Labs Research, Meta}
  \city{Redmond}
  \state{WA}
  \country{USA}}
\email{yanx@meta.com}

\author{Raj Sodhi}
\affiliation{%
  \institution{Reality Labs Research, Meta}
  \city{Redmond}
  \state{WA}
  \country{USA}}
\email{rsodhi@meta.com}

\renewcommand{\shortauthors}{Wang et al.}

\definecolor{myblue}{RGB}{110, 245, 227}
\newcommand{\yuliang}[1]{{{\textcolor{myblue}{\{Yuliang: \bf #1\}}}\xspace}}

\begin{abstract}

Video creation has become increasingly popular, yet the expertise and effort required for editing often pose barriers to beginners. In this paper, we explore the integration of large language models (LLMs) into the video editing workflow to reduce these barriers. Our design vision is embodied in LAVE, a novel system that provides LLM-powered agent assistance and language-augmented editing features. LAVE automatically generates language descriptions for the user's footage, serving as the foundation for enabling the LLM to process videos and assist in editing tasks. When the user provides editing objectives, the agent plans and executes relevant actions to fulfill them. Moreover, LAVE allows users to edit videos through either the agent or direct UI manipulation, providing flexibility and enabling manual refinement of agent actions. Our user study, which included eight participants ranging from novices to proficient editors, demonstrated LAVE's effectiveness. The results also shed light on user perceptions of the proposed LLM-assisted editing paradigm and its impact on users' creativity and sense of co-creation. Based on these findings, we propose design implications to inform the future development of agent-assisted content editing.

\end{abstract}
\keywords{Video Editing, LLMs, Agents, Human-AI Co-Creation}
\begin{teaserfigure}
  \includegraphics[width=\textwidth]{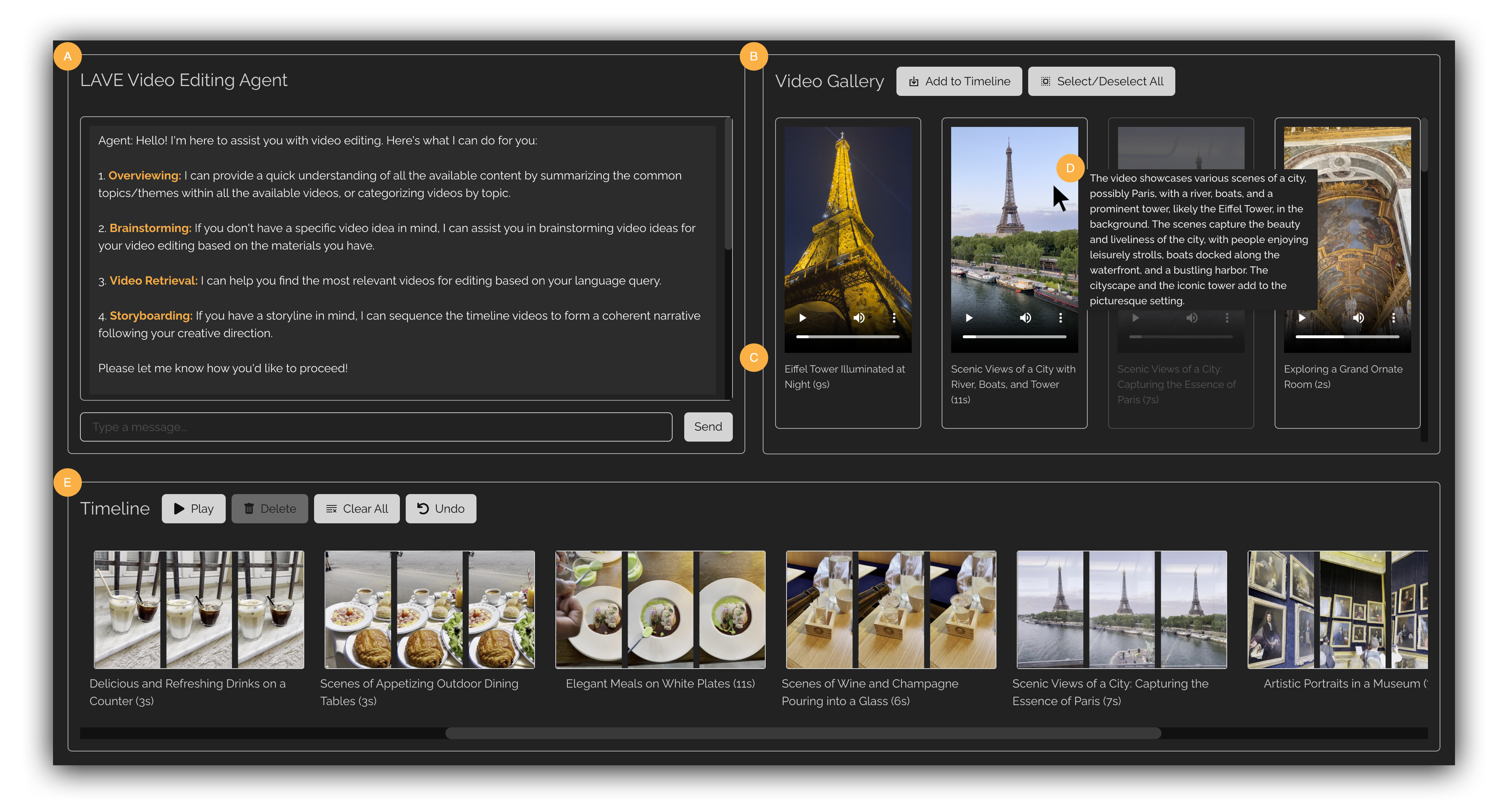}
  \caption{The LAVE system is a video editing tool that offers LLM-powered agent assistance and language-augmented features. A) LAVE's video editing agent assists with several video editing tasks, with which users can converse to obtain agent assistance throughout the editing process. B) A language-augmented video gallery. Users can click on a desired video to select and add it to the editing timeline. Videos added to the timeline will be displayed in reduced opacity. C) LAVE automatically generates succinct titles for each video. D) Hovering over a video in the gallery displays a tooltip with the video summary, allowing users to understand the video content without playing it. E) An editing timeline where users can reorder and trim clips. These edits can be performed either with LLM assistance or manually.}
  \Description{.}
  \label{fig:teaser}
\end{teaserfigure}

\maketitle
\section{Introduction}
Videos are a powerful medium for communication and storytelling. Their popularity has surged with the advent of social media and video-sharing platforms, inspiring many to produce and share their content. However, the complexity of video editing can pose significant barriers for beginners. For example, the initial ideation and planning phases, crucial in the early stages of the creative process, can be challenging for those unfamiliar with video concept development. Furthermore, editing operations often involve meticulous selection, trimming, and sequencing of clips to create a coherent narrative. This not only requires mastery of the often complex user interfaces of editing software but also significant manual effort and storytelling skills.

Recently, natural language has been used to address the challenges associated with video editing. Utilizing language as an interaction medium for video editing allows users to directly convey their intentions, bypassing the need to translate thoughts into manual operations.
For instance, recent AI products \cite{runway} allow users to edit video leveraging the power of text-to-video models \cite{singer2022makeavideo, ho2022imagen}; voice-based video navigation enables users to browse videos using voice commands instead of manual scrubbing \cite{10.1145/3411764.3445131,10.1145/3290605.3300931}. In addition, language has been used to represent video content, thereby streamlining the manual editing process. A prominent example is text-based editing, which enables users to efficiently edit a narrative video by adjusting its time-aligned transcripts \cite{ohad2019, 10.1145/3379337.3415864, huber2019b, 10.1145/3544548.3581494}. Despite these advancements, the majority of video editing tools still heavily rely on manual editing and often lack customized, in-context assistance. Consequently, users are left to grapple with the intricacies of video editing on their own.\\

\textbf{How can we design a video editing tool that acts as a collaborator, constantly assisting users in the editing process?} Such a tool could help users generate video editing ideas, browse and find relevant clips, and sequence them to craft a compelling narrative. Building upon previous work that integrates natural language with video editing, we propose to instrument video editing with LLM's versatile linguistic capabilities, e.g., storytelling and reasoning, which have proven useful in assisting various creative tasks \cite{10.1145/3490099.3511105, 10.1145/3491102.3501819, 10.1145/3544548.3581225, chakrabarty2023creativity, liu2022opal, liu20233dall, wang2023reelframer, liu2023generative}. In doing so, we probe into a future video editing paradigm that, through the power of natural language, reduces the barriers typically associated with manual video editing.

We present LAVE, a video editing tool that offers language augmentation powered by LLMs. LAVE introduces an LLM-based plan-and-execute agent capable of interpreting users' free-form language commands, planning, and executing relevant actions to achieve users' editing objectives. These actions encompass conceptualization assistance, such as brainstorming ideas and summarizing a video corpus with an overview, as well as operational assistance, including semantic-based video retrieval, storyboarding (sequencing videos to form a narrative), and trimming clips. To enable these agent actions, LAVE automatically generates language descriptions of the video's visuals using visual-language models (VLMs). These descriptions, which we refer to as visual narrations, allow LLMs to understand the video content and leverage their linguistic capabilities to assist users in editing tasks. LAVE offers two interaction modalities for video editing: agent assistance and direct manipulation. The dual modalities provide users with flexibility and allow them to refine agent actions as needed.

We conducted a user study with eight participants, which included both novice and proficient video editors, to assess the effectiveness of LAVE in aiding video editing. The results demonstrated that participants could produce satisfactory AI-collaborative video outcomes using LAVE. Users expressed appreciation for the system's functionalities, finding them easy to use and useful for producing creative video artifacts. Furthermore, our study uncovered insights into users' perceptions of the proposed editing paradigm, their acceptance of agent assistance across different tasks, as well as the system's influence on their creativity and sense of human-AI co-creation. Based on these findings, we proposed design implications to inform the development of future multimedia content editing tools that integrate LLMs and agents. In summary, this paper makes the following contributions: 

\begin{itemize}
    \item The conceptualization and implementation of the LAVE system, a language-augmented video editing tool that leverages LLM's linguistic intelligence to facilitate an agent-assisted video editing experience. \\
    
    \item The design of an LLM-based computational pipeline that enables LAVE's video editing agent to plan and execute a range of editing functions to help achieve users' editing objectives.\\

    \item The user study results showcasing the advantages and challenges of integrating LLMs with video editing. The findings highlight user perceptions and emerging behaviors with the proposed editing paradigm, from which we propose design implications for future agent-assisted content editing.

\end{itemize}

\section{Related Work}
LAVE builds upon existing work in language as a medium for video editing, LLM and agents, and human-AI co-creation.
\subsection{Language as Medium for Video Editing} 

Traditional video editing tools like Premier Pro \cite{premierpro} and Final Cut Pro \cite{finalcut} demand manual interaction with raw clips. While precise, it can be cumbersome due to UI complexity. Additionally, visual elements of raw footage such as thumbnails and audio waveforms might not always convey its semantics effectively. Language, on the other hand, offers an intuitive and efficient alternative to complex UI in video editing and has been investigated in video editing tool research \cite{xia2020crosscast, crosspower, ohad2019, 10.1145/3379337.3415864, huber2019b, 10.1145/3544548.3581494, quickcut}.  One common approach treats language as a "\textit{Command}", where users employ language to instruct tools for specific operations. This is evident in multimodal authoring tools that support speech commands \cite{pixeltone} and voice-based video navigation \cite{lin2023identifying, 10.1145/3290605.3300931}. However, existing work primarily supports single-turn interactions and provides a limited range of commands. As a result, they do not accommodate diverse language and long-term conversations. In contrast, LAVE accepts free-form language, supporting natural interaction and allowing back-and-forth discussions with an agent throughout the video editing process. Another significant body of work treats language as "\textit{Content}", where language becomes part of the content being edited. For instance, text-based editing for narrative videos \cite{ohad2019, 10.1145/3379337.3415864, huber2019b, 10.1145/3544548.3581494} and creating video montages by scripting \cite{wang2019write}. Nevertheless, these techniques rely on either the pre-existing language content in the videos, such as narration, or on language annotations provided by the user \cite{quickcut, wang2019write}. The former is often missing in everyday videos recorded by individuals, while the latter requires additional manual effort. In contrast, LAVE automatically generates language descriptions for each video and leverages LLM's linguistic capabilities to automate and facilitate content editing. Recent work in generative AI, such as Make-A-Video \cite{singer2022makeavideo} and Imagen Video \cite{ho2022imagen}, have investigated synthesizing videos from textual prompts using diffusion techniques. Unlike these efforts, which aim to generate new footage, our objective is to facilitate the editing of existing videos. That said, we anticipate that video generation techniques will complement editing tools like LAVE, especially in use cases like creating B-rolls.

\subsection{Large Language Models and Agents}

LLMs, such as GPT-4 \cite{openai2023gpt4} and LLaMA \cite{touvron2023llama}, are trained on vast amounts of text data and possess immense model sizes. They have been shown to encode a wealth of human knowledge \cite{huang2022language, roberts-etal-2020-much, li2021implicit} and can perform sophisticated reasoning \cite{wei2023chainofthought,  kojima2023large, nye2021work} and action planning \cite{huang2022language}. Their linguistic and storytelling capabilities have been utilized in creative writing \cite{10.1145/3490099.3511105, 10.1145/3491102.3501819, 10.1145/3544548.3581225, chakrabarty2023creativity} and a myriad of other creative applications \cite{liu2022opal, liu20233dall,  wang2023reelframer, liu2023generative, brade2023promptify}. Moreover, LLMs can adapt to new tasks based on a given description without re-training, a method known as \textit{prompting}. Owing to the efficiency and adaptability, there has been a surge in interest in prompting techniques \cite{zamfirescu2023johnny, kim2023evallm, arawjo2023chainforge, brown2020language, 10.1145/3544548.3580895, logan2021cutting}. Notable ones include few-shot prompting \cite{brown2020language}, where multiple input/output data examples are provided to enhance task performances, and chain-of-thought prompting \cite{wei2023chainofthought}, which directs the LLM in generating a sequence of intermediate reasoning steps prior to the final output. Leveraging these techniques, recent studies have explored the development of agents autonomously interacting with various environments using LLMs \cite{wang2023planandsolve, park2023generative, shaw2023pixels, song2023llm, bran2023chemcrow, shinn2023reflexion, yao2023react, li2023zeroshot}. For example, Wang et al. \cite{wang2023planandsolve} introduced an agent that devises a plan dividing tasks into subtasks and executes them. Yao et al. \cite{yao2023react} presented the ReAct framework, where LLMs generate interleaved reasoning sequences and task-specific actions. This paper builds upon prior work in this area and proposes an agent architecture designed for interactive video editing, which plans and executes relevant editing actions based on the user's instructions.

\subsection{Human-AI Co-Creation}
As AI continues to advance in its capability to generate content and automate tasks, it is being increasingly incorporated into the creative processes across various domains \cite{10.1145/3544548.3581225,  10.1145/3490099.3511105, 10.1145/3491102.3501819, 10.1145/3544548.3581225, chakrabarty2023creativity, 10.1145/3490099.3511159, huang2020ai, 10.1145/3313831.3376739, 10.1145/3526113.3545617, rope}. This includes areas such as story writing \cite{10.1145/3491102.3501819, 10.1145/3490099.3511105, 10.1145/3544548.3581225, chakrabarty2023creativity}, music composition \cite{10.1145/3490099.3511159, huang2020ai, 10.1145/3313831.3376739}, comic creation \cite{10.1145/3526113.3545617}, and game design \cite{8490433}. For instance, TaleBrush \cite{10.1145/3491102.3501819} enables users to craft stories with the support of language models by sketching storylines metaphorically. Storybuddy \cite{zhang2022storybuddy} produces interactive storytelling experiences by generating story-related questions. Cococo \cite{10.1145/3313831.3376739} investigates the challenges and opportunities inherent in co-creating music with AI, especially for beginners. CodeToon \cite{10.1145/3526113.3545617} automatically converts code into comics. However, while AI holds significant promise for enhancing a user's creative abilities by managing certain aspects of the creative workflow, it also brings forward challenges and concerns such as user agency and trusts \cite{kang2022ai}, the authenticity of the creation \cite{mccormack2019autonomy}, potential creative biases \cite{magni2023humans, loughran2022bias}, and ownership and credit attribution \cite{eshraghian2020human, bisoyi2022ownership}. Our work builds upon existing literature in human-AI co-creation  \cite{10.1145/3359313, rezwana2022identifying, buschek2021nine, park2019identifying, khadpe2020conceptual, amershi2019guidelines, liu2021ai, glikson2020human, eshraghian2020human, kang2022ai} and further contributes by developing a new AI system for video editing and studying its impact. Through the lens of LAVE, we examined the dynamics of user interactions with an LLM-based agent and explored the opportunities and challenges inherent in the proposed editing paradigm.

\section{Design Goals}
This work aims to explore the potential of a collaborative experience between humans and LLM agents in video editing through the design, implementation, and evaluation of the LAVE system. To this end, we outlined two primary design goals that serve as the guiding principles for the system design.
\\

\noindent \textbf{D1. Harnessing Natural Language to Lower Editing Barriers.}
The central proposition of this work is to enhance manual video editing paradigms with the power of natural language and LLMs. We intended to design LAVE to lower barriers to editing for users by leveraging the linguistic intelligence of LLMs from the initial ideation to the editing operations. \\

\noindent \textbf{D2. Preserving User Agency in the Editing Process.}
A common concern regarding AI-assisted content editing is the potential loss of user autonomy and control. To mitigate this concern, we designed LAVE to offer both AI-assisted and manual editing options. This allows users to refine or opt out of AI assistance as needed, thereby preserving user agency. It ensures that the final product reflects the user's artistic vision and grants them decision-making authority.

\begin{figure*}[h]
    \centering
    \includegraphics[width=1\linewidth]{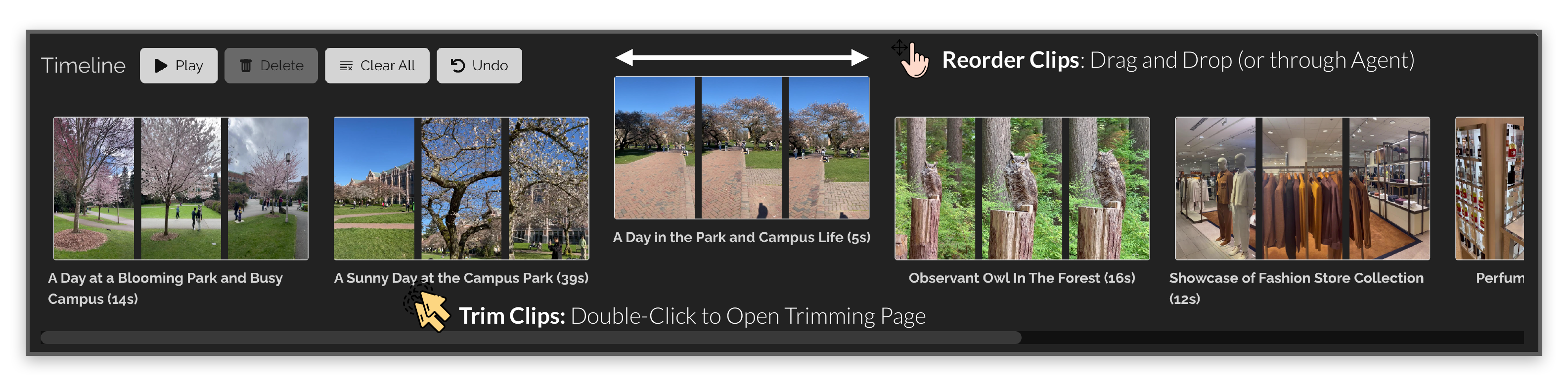}
    \caption{LAVE's video editing timeline: Users can drag and drop video clips to rearrange their order. The order can also be changed through LAVE's video editing agent's storyboarding function. To trim a clip, users can double-click it, revealing a pop-up window for trimming as shown in Figure \ref{fig:trimming} .}
    \label{fig:timeline}
\end{figure*}

\section{The LAVE User Interface}
\label{laveui}
Guided by the design goals, we developed the LAVE system. LAVE's UI comprises three primary components: 1) the Language Augmented Video Gallery, which displays video footage with automatically generated language descriptions; 2) the Video Editing Timeline, containing the master timeline for editing; and 3) the Video Editing Agent, enabling users to interact with a  conversational agent and receive assistance. When users communicate with the agent, the message exchanges are displayed in the chat UI. The agent can also make changes to the video gallery and the editing timeline when relevant actions are taken. Additionally, users can interact directly with the gallery and timeline using a cursor, similar to traditional editing interfaces. In the subsequent sections, we describe the details of each component and highlight their connection to the design goals.

\subsection{Language-Augmented Video Gallery}
\label{gallery}
\begin{figure}[h]
    \centering
    \includegraphics[width=1\linewidth]{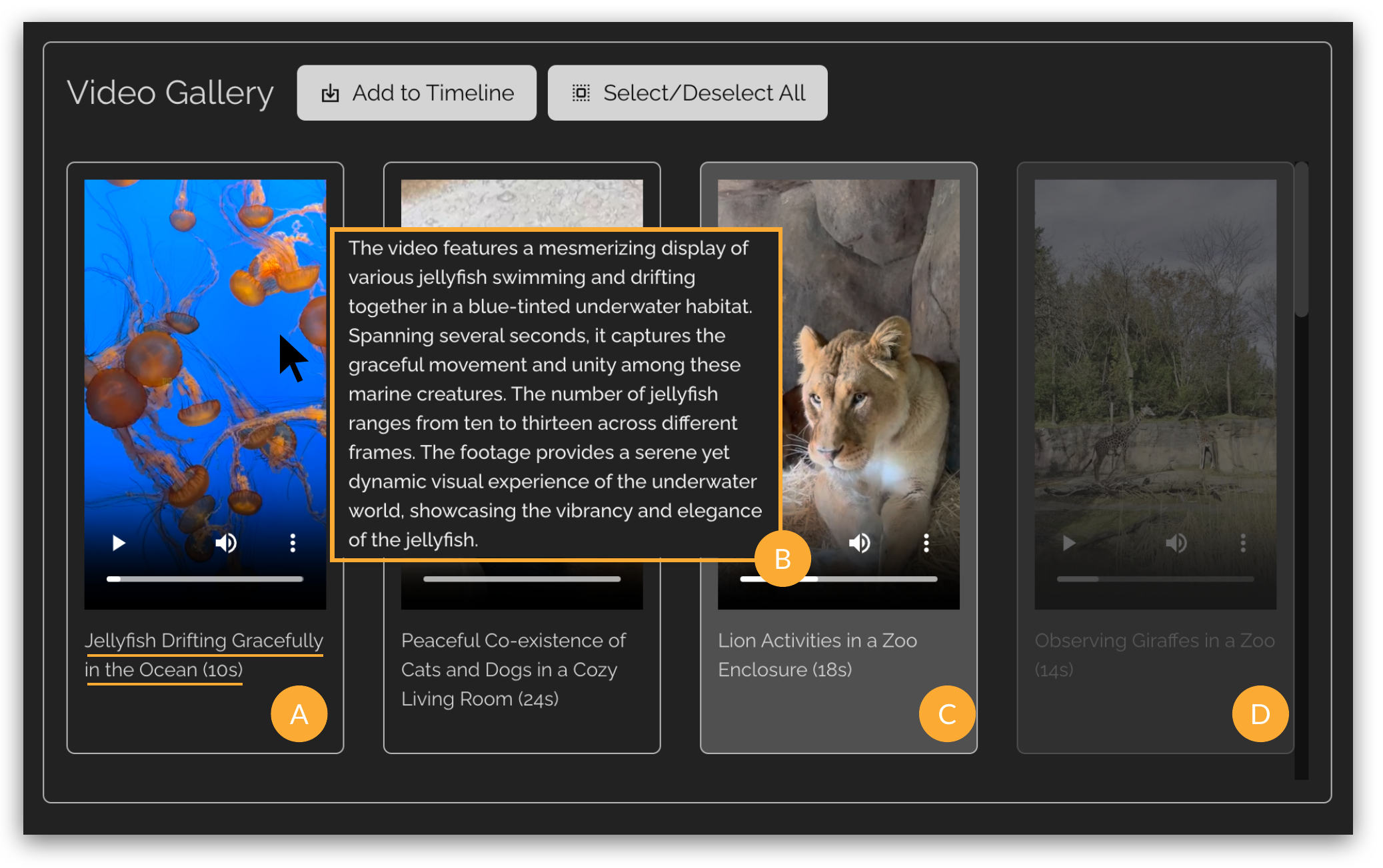}
    \caption{LAVE's language-augmented video gallery features each video with a semantic title and its length (A). When users hover their cursor over a video, a detailed summary appears, allowing them to preview the video content without playing it (B). Users can select multiple videos to add to the timeline.  Selected videos will be highlighted in light grey (C) and those already added will appear with faded opacity (D). }
    \label{fig:gallery}
\end{figure}

LAVE features a language-augmented video gallery, as shown in Figure \ref{fig:gallery}. Like traditional tools, it allows clip playback but uniquely offers visual narrations, i.e., auto-generated textual descriptions for each video, including semantic titles and summaries. The titles can assist in understanding and indexing clips without needing playback. The summaries provide an overview of each clip's visual content, which could assist users in shaping the storylines for their editing projects.  The title and duration are displayed under each video. Hovering over a video reveals a tooltip with the narrative summary. Users can select clips to add to the editing timeline using the `Add to Timeline' button. If users wish to use all of their videos (e.g., all footage from a trip), they can simply use the \textit{`Select/Deselect All'} option to add them to the timeline. Moreover, LAVE enables users to search for videos using semantic language queries, with the retrieved videos presented in the gallery and sorted by relevance. This function must be performed through the editing agent, which we will discuss further in the corresponding sections.

\subsection{Video Editing Timeline}
\label{timeline}
Once videos are selected from the video gallery and added to the editing timeline, they are displayed on the video editing timeline at the bottom of the interface (Figure \ref{fig:timeline}). Each clip on the timeline is represented by a box that showcases three thumbnails: the start, midpoint, and end frames of the video to illustrate its content. In the LAVE system, each thumbnail frame represents one second worth of footage within the clip. As in the video gallery, the titles and descriptions of each clip are also provided. The editing timeline in LAVE features two key functions: clip sequencing and trimming. Each offers LLM-based and manual options, affording users flexibility and control over AI assistance \textbf{(D2)}.

\subsubsection{Clip Sequencing}
Sequencing clips on a timeline is a common task in video editing, essential for creating a cohesive narrative. LAVE supports two sequencing methods: 1) LLM-based sequencing operates via the storyboarding function of LAVE's video editing agent. This function orders clips based on a user-provided or LLM-generated storyline. We will further this feature in the agent sections. 2) Manual sequencing allows users to arrange clips through direct manipulation, enabling them to drag and drop each video box to set the order in which the clips will appear. If users want to remove videos from the timeline, they can select specific clips and click the \textit{"Delete"} button. There is also a \textit{"Clear All"} option for removing all videos from the timeline simultaneously. Additionally, users can reverse any edits using the \textit{"Undo"} button. To preview the combined output of the current clip sequence, users can click the \textit{"Play"} button, after which the system generates a preview video for review.

\subsubsection{Clip Trimming}

\begin{figure}[t]
    \centering
    \includegraphics[width=1\linewidth]{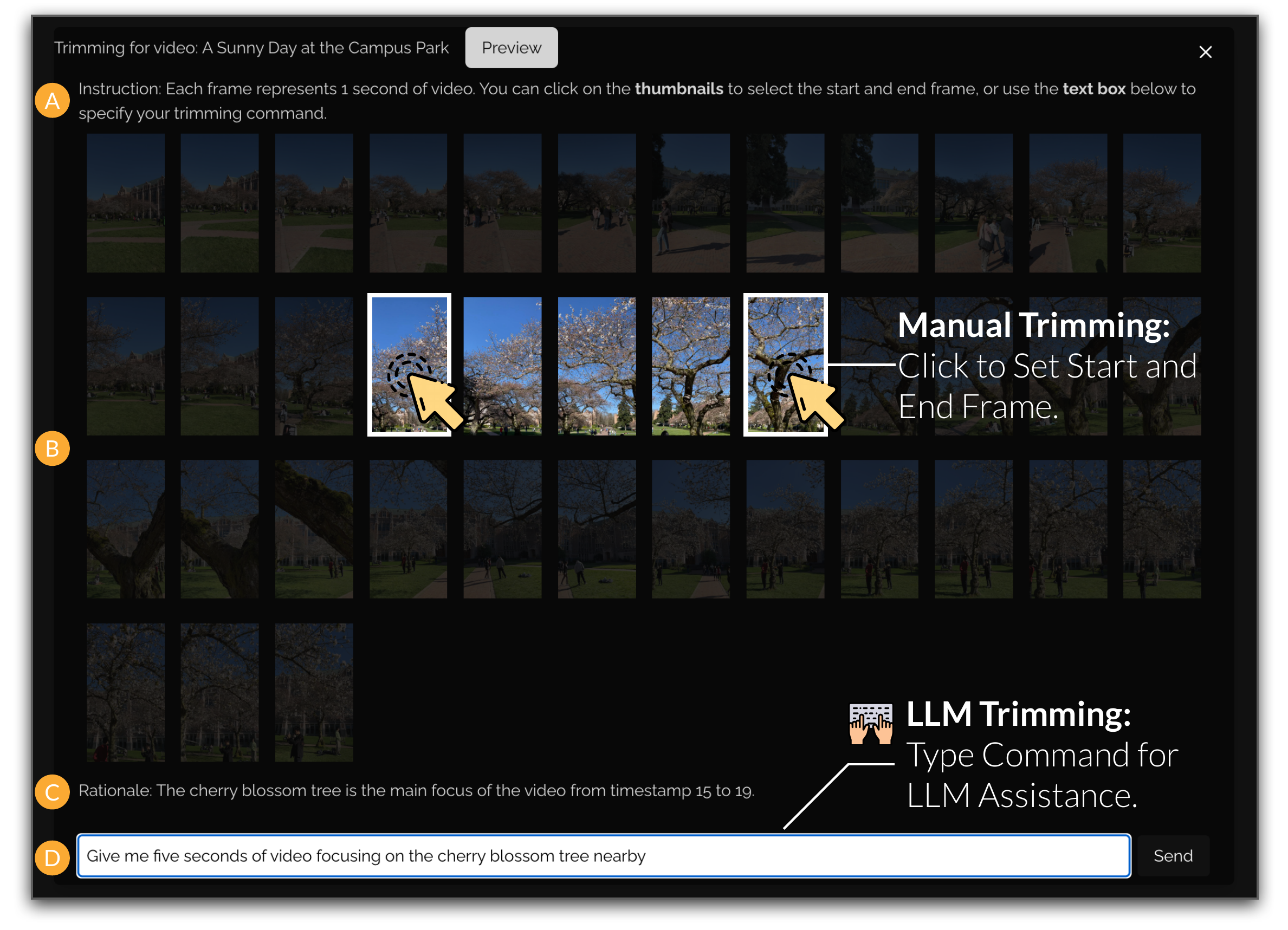}
    \caption{LAVE's clip-trimming window displays user guide (A) and video frames sampled every second from the clip (B). Users can manually set the start and end frames for trimming. Alternatively, they can use the LLM-powered trimming feature with commands like \textit{"Give me 5 seconds focusing on the nearby cherry blossom tree."} (D). With this approach, the trim automatically adjusts and includes a rationale explaining the LLM's choice (C). Frames not included in the trimmed clip are displayed in a dimmed color.}

    \label{fig:trimming}
\end{figure}

Trimming is essential in video editing to highlight key segments and remove redundant content. To trim, users double-click a clip in the timeline, opening a pop-up that displays one-second frames (Figure \ref{fig:trimming}). Similar to Clip sequencing, LAVE supports both LLM-based and manual clip trimming:
\begin{itemize}
    \item \textbf{LLM-based Trimming}: Below the frames, a text box is provided for users to input trimming commands to extract video segments based on their specifications. These commands can be free-form. For instance, they might refer to the video's semantic content, such as \textit{"keep only the segment focusing on the baseball game"}, or specify precise trimming details like \textit{"Give me the last 5 seconds."} Commands can also combine both elements, like \textit{"get 3 seconds where the dog sits on the chair"}. This functionality harnesses the LLM's information extraction capability \cite{agrawal2022large} to identify segments aligning with user descriptions. For transparency, the LLM also explains its rationale for the trimmings, detailing how they align with user instructions. Note that while the feature is also powered by LLM, it is not part of the LAVE editing agent's operations, which primarily handle video operations at the project level. This trimming feature is specifically designed for individual clip adjustments.
    
    \item \textbf{Manual Trimming}: Users can manually select frames to define the starting and ending points of a clip by clicking on the thumbnails. This feature also allows users to refine LLM-based trimming when it does not align with their intentions.
\end{itemize}

\begin{figure*}[t]
    \centering
    \includegraphics[width=\linewidth]{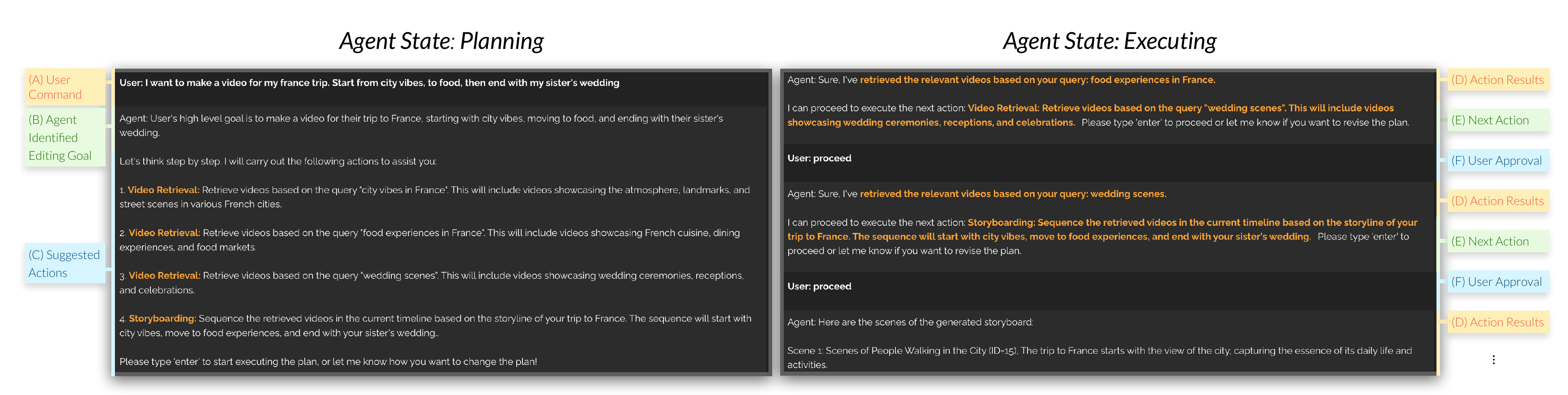}
    \caption{LAVE's video editing agent operates in two states: Planning and Executing. In the Planning state (left), users provide editing commands (A). The agent then clarifies the goal (B) and proposes actionable steps to achieve the goal (C). Users have the option to revise the plan if they are not satisfied with the proposed steps. Upon user approval of the plan, the agent transitions to the Executing state (right). In this state, the user approves the agent's actions sequentially. Following each action, the agent presents the results (Ds). If additional actions are outlined in the plan, the agent notifies the user of the next action (Es) and waits for their approval (Fs).}
\label{fig:agent-chat}
\end{figure*}

\subsection{Video Editing Agent}
\label{lave_agent}
LAVE's video editing agent is a chat-based component that facilitates interactions between the user and an LLM-based agent. Unlike command-line tools, users can interact with the agent using free-form language. The agent offers video editing assistance leveraging the linguistic intelligence of LLMs and can provide tailored responses to guide and assist users throughout the editing process \textbf{(D1)}.
LAVE's agent assistance is provided through agent actions, each involving the execution of an editing function supported by the system. In the following sections, we outline the interaction experience with the agent and describe the editing functions.

\subsubsection{Interacting with the Plan-and-Execute Agent}
\label{agentinteract}
To collaborate with the agent, users begin the process by typing their editing objectives. The agent interprets the user's objectives and formulates an action plan to fulfill them \cite{karpas2022mrkl, yao2023react, shinn2023reflexion, shen2023hugginggpt}.  The agent operates in two modes: \textit{Planning} and \textit{Executing}. By default, the agent starts in the \textit{Planning} state (Figure \ref{fig:agent-chat}-left). In this state, whenever a user inputs an editing goal, the agent evaluates it to determine what actions to perform to fulfill the user's goal. The agent can execute multiple actions, particularly when a user's objective is broad and involves diverse operations. For instance, if a user types, \textit{"I want to make a video but I don't have any ideas,"} the agent may propose a plan that includes brainstorming ideas, finding relevant footage, and arranging clips to craft a narrative based on the brainstormed concepts. On the other hand, users can also issue a specific command so the action plan contains exactly one desired action. The proposed plan requires user approval before execution and the user can request adjustments or clarifications \textbf{(D2)}. 

Execution begins after the user presses "enter"—this user approval transitions the agent to the \textit{Executing} state, wherein it begins executing the planned actions sequentially (Figure \ref{fig:agent-chat}-right). After each action is carried out, the agent informs the user of the results and the next action, if available. The user can then either press "enter" again to proceed with subsequent actions or engage with the agent to alter or cancel the remaining plan. The agent maintains a memory buffer for previous conversations, allowing it to access the recent context when proposing functions. For example, if the agent has previously brainstormed ideas with the user, it might suggest performing video retrieval based on the idea the user selected. 

\subsubsection{Editing Functions}
\label{agent_operations}
LAVE's agent supports four editing functions: Footage Overviewing and Idea Brainstorming provide conceptualization assistance based on LLM's summarization and ideation abilities, respectively. The other two, Video Retrieval and Storyboarding, leverage LLM's embedding and storytelling capabilities, respectively, to facilitate the manual editing process.

\begin{itemize}
    \item \textbf{Footage Overviewing:} The agent can generate an overview text that summarizes the videos the user provided in the gallery, categorizing them based on themes or topics. For instance, clips from a road trip to the Grand Canyon might be categorized under themes like "Hiking and Outdoor Adventures" or "Driving on Highways." This feature is particularly helpful when users are not familiar with the footage, such as when editing videos from older collections or dealing with extensive video sets.\\

    \item \textbf{Idea Brainstorming:}
    The agent can assist in brainstorming video editing ideas based on the gallery videos. This allows the agent to suggest various concepts, helping to ignite the users' creative sparks. For example, the agent might suggest using several clips of the user's pet to create a video on the topic, "A Day in the Life of Pets—from Day to Night." Additionally, users can provide the agent with optional creative guidance or constraints to guide the agent's ideation process. \\
    
    \item \textbf{Video Retrieval:} Searching for relevant footage is a fundamental yet often tedious aspect of video editing. Instead of the user manually searching the gallery, the agent can assist by retrieving videos based on language queries, such as "Strolling around the Eiffel Tower." After completing the retrieval, the agent will present the most relevant videos in the language-augmented video gallery, sorted by relevance. \\

    \item \textbf{Storyboarding:} Video editing often requires sequencing clips in the timeline to construct a specific narrative. The agent can assist users in ordering these clips based on a narrative or storyline provided by the users. The narrative can be as concise as \textit{"from indoor to outdoor"}, or more detailed, for example, \textit{"starting with city landscapes, transitioning to food and drinks, then moving to the night social gathering."} If users do not provide a storyline, the agent will automatically generate one based on the videos already added to the timeline. Once the agent generates a storyboard, the videos in the timeline will be re-ordered accordingly. The agent will also provide a scene-by-scene description of the storyboard in the chatroom.
    
\end{itemize}

\subsection{Supported Workflows and Target Use Cases}
Altogether, LAVE provides features that span a workflow from ideation and pre-planning to the actual editing operations. However, the system does not impose a strict workflow. Users have the flexibility to utilize a subset of features that align with their editing objectives. For instance, a user with a clear editing vision and a well-defined storyline might bypass the ideation phase and dive directly into editing. In addition, LAVE is currently designed for casual editing, such as creating videos for social media platforms. We leave the integration of LLM agents into professional editing, where utmost precision is crucial, as future work.

\begin{figure*}[!t]
    \centering
    \includegraphics[width=1\linewidth]{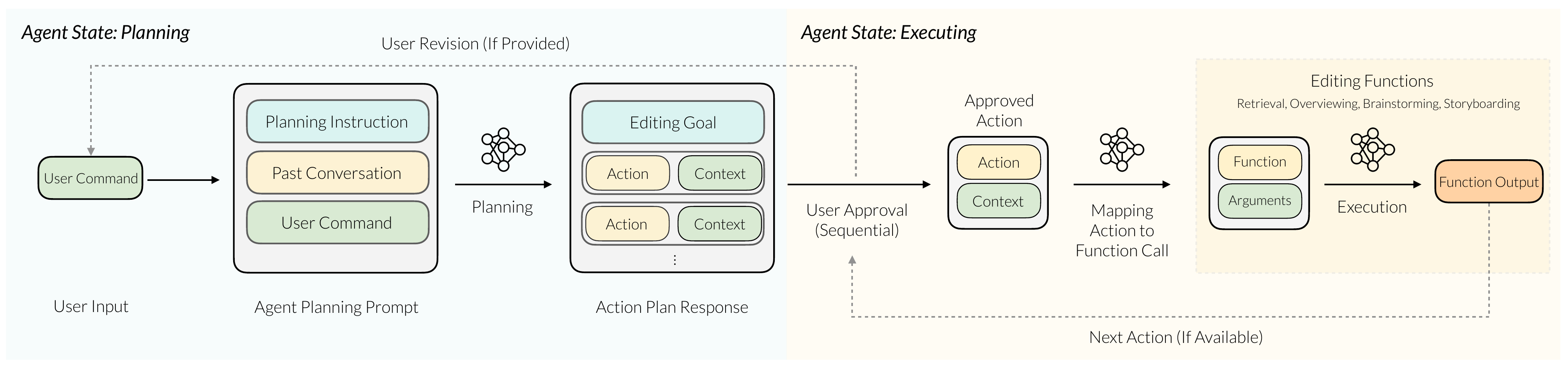}
    \caption{LAVE's plan-and-execute agent design: Upon receiving an input containing the user's editing command, a planning prompt is constructed. This prompt includes the planning instruction, past conversations, and the new user command. It is then sent to the LLM to produce an action plan, which reflects the user's editing goal and outlines actions to assist the user in achieving this goal. Each action is accompanied by a context, which provides additional information relevant to the action, such as a language query for video retrieval. The user reviews and approves the actions one by one. After an action is approved, it is translated into actual Python function calls and executed. This process continues for all the actions in the plan, unless the user decides to provide new instructions to revise or cancel the plan.}
    
   \label{fig:agent-design}
\end{figure*}

\section{Backend System}
We now describe the backend processing and system design that enable the interactive components outlined in Section \ref{laveui}. We start by describing the design of LAVE's video editing agent and delve deeper into the implementation of the editing functions. We utilize OpenAI's GPT-4 \cite{openai2023gpt4} for all LLM mentions in the subsequent sections unless stated otherwise.

\subsection{Agent Design}
We built the LAVE agent by leveraging LLMs' diverse language capabilities, including reasoning, planning, and storytelling. The LAVE Agent has two states: Planning and Executing. The plan-and-execute approach offers two primary benefits: 1) It allows users to set high-level objectives encompassing multiple actions, removing the necessity to detail every individual action like traditional command line tools. 2) Before execution, the agent presents the plan to the user, providing a chance for revisions and ensuring that users maintain complete control \textbf{(D2)}. We designed a backend pipeline to facilitate this plan-and-execute agent.  As depicted in Figure \ref{fig:agent-design}, the pipeline begins by creating an action plan based on user input. This plan is then translated from textual descriptions into function calls, which subsequently execute the corresponding functions. We expand on the specifics of each step in the subsequent sections.

\subsubsection{Action Planning}
The action planning of LAVE's video editing agent employs a specialized LLM prompt format, which is informed by previous research on LLM prompting. We incorporated action/tool-use agent prompting techniques \cite{karpas2022mrkl, yao2023react, shinn2023reflexion, shen2023hugginggpt}. In this context, the "tools" or "actions" are equal to the system's editing functions. We also leveraged insights from the chain-of-thought prompting \cite{wei2023chainofthought}, which uses LLMs' reasoning capabilities to decompose complex tasks (user goals) into sub-tasks (editing functions). The prompt preamble of our system consists of three segments.
\begin{enumerate}
    \item \textbf{Role Assignment}: An opening paragraph directing the agent to act as a video editing assistant tasked with generating an action plan from user commands.
    \item \textbf{Action Descriptions}: Following the role assignment, we describe a list of actions that the agent can perform. Each action corresponds to an editing function supported by LAVE. We detail the functionality and use cases of each, assisting the agent in selecting appropriate responses to meet the user's commands.
    \item \textbf{Format Instruction}: Lastly, we guide the agent to output the action plan in a consistent format: First, determine the user's editing goal, followed by a stepwise plan enumerating suggested actions to achieve that goal. Each action includes the function name and its associated context, if applicable. For instance, \textit{"Storyboarding (function name): Create a storyboard from day to night. (context)"} We also instruct the model to prefix the user's goal and action list with the capitalized words "GOAL" and "ACTIONS," respectively, to facilitate text parsing for downstream processing.

\end{enumerate}
 
After the preamble, we append the recent conversation history, along with the latest user input. This combination forms the complete prompt sent to the LLM for generating an action plan. The conversation history is useful when the user wants to refer to a previous message or a generated plan, e.g., if they want to change a plan that the agent proposed. The system retains up to 6000 tokens of message history. If this limit is exceeded, it will begin removing messages starting with the second oldest, while preserving the oldest message, i.e., the preamble. The 6000-token limit, set empirically, is approximately 2000 tokens fewer than the context window of the LLM used, ensuring space for text generation (25\% of context limit). This setting can be adjusted to accommodate the lengths of different LLMs' context windows. The tokens are byte pair encoding (BPE) \cite{shibata1999byte} tokens utilized by LLMs such as GPT-4.

\subsubsection{Translating Action Plan to Executable Functions}
\label{mapping}
As discussed in Section \ref{agentinteract}, upon formulating an action plan, it is presented to the user for approval. Rather than batch approval, each action is sequentially approved by the user. This method allows the user to execute one action, observe its results, and then decide whether to proceed with the subsequent action. To facilitate this process, LAVE parses each action description from the action plan and translates it into a corresponding backend function call. We utilize an OpenAI GPT-4 checkpoint, which has been fine-tuned for Function Calling \cite{functioncalling}, to accomplish this translation. To make use of the Function Calling feature, we provide detailed descriptions of each function's capabilities. Once completed, the LLM can transform a textual prompt, specifically an action description in our case, into the corresponding editing function call with contextually extracted arguments. The results of the function execution are updated in the frontend UI and presented to the user.

\begin{table*}
    \centering
    \caption{Input, output, and the parts of the UI that receive updates for each LLM-powred editing function. Gallery Videos and Timeline Videos refer to the visual narration of the corresponding videos in text format. Optional Guidance indicates that the user can provide extra, optional input to guide the function.}

    \label{functionio}
    \begin{tabular}{llll}
        \toprule
        Function & Input & Output & UI Updates\\
        \midrule
        Video Retrieval & Text Query + Vector Store & Ranked Videos & Video Gallery \\
        Footage Overviewing & Gallery Videos  & Overview & Agent Chat  \\
        Idea Brainstorming & Gallery Videos  + Optional Guidance& Ideas & Agent Chat  \\
        Storyboarding  & Timeline Videos + Optional Guidance  & Storyboard + Video Order  & Agent Chat + Timeline\\
        Clip Trimming  & Frame Captions + Trimming Command  & Start/End Frame IDs + Rationale  & Timeline\\
        \bottomrule
    \end{tabular}
\end{table*}

\subsection{Implementation of LLM-Powered Editing Functions}
LAVE supports five LLM-powered functions to assist users in video editing tasks: 1) Footage Overview, 2) Idea Brainstorming, 3) Video Retrieval, 4) Storyboarding, and 5) Clip Trimming. The first four of them are accessible through the agent (Figure \ref{fig:agent-chat}), while clip trimming is available via the window that appears when double-clicking clips on the editing timeline (Figure \ref{fig:trimming}). Among them, language-based video retrieval is implemented with a vector store database, while the rest are achieved through LLM prompt engineering. All functions are built on top of the automatically generated language descriptions of raw footage, including the titles and summaries of each clip as illustrated in the video gallery (Figure \ref{fig:gallery}). We refer to these textual descriptions of videos as \textit{visual narrations} as they describe the narratives in the visual aspects of the video. Table \ref{functionio} outlines the input, output, and UI updates for each function. Figure \ref{function-illustrations} in the appendix provides additional illustrations of each function's mechanism. In the following sub-sections, we start by describing the pre-processing that generates visual narrations and then delve into the implementation of each function. 

\subsubsection{Generating Visual Narration: Video Title and Summary}
\label{preproceessing}
The process of generating visual narrations involves sampling video frames at a rate of one frame per second. Each frame is then captioned using LLaVA \cite{liu2023llava} v1.0, which is built upon Vicuna-V1-13B \cite{vicuna2023}, a fine-tuned checkpoint of LLaMA-V1-13B model \cite{touvron2023llama}. After compiling the frame descriptions, we leverage GPT-4 to generate titles and summaries. Furthermore, each video is assigned a unique numeric ID. This ID aids the LLM in referencing individual clips for functions such as storyboarding. Note that during the development phase, we chose LLaVA due to its language model's ability to generate more comprehensive captions than other contemporary models commonly used for image captioning, such as BLIP-2 \cite{li2023blip2}. However, we were aware of the rapid evolution of VLMs, and that newer models might soon outperform LLaVA v.1.0. We discuss the integration of these models in the future work section.

\subsubsection{Video Retrieval based on Text Embedding}
LAVE's video retrieval feature utilizes a vector store, constructed by embedding the visual narrations (titles and summaries) of each video using OpenAI's \texttt{text-embedding-ada-002}. This process results in 1536-dimensional embeddings for each video. During retrieval, LAVE embeds the query, identified from the user's command, with the same model and computes the cosine distances between the query and the stored video embeddings to rank the videos accordingly. Subsequently, LAVE updates the frontend UI video gallery with videos sorted based on the ranking. Although our design primarily focuses on a ranking-based approach for displaying the retrieved results, it can easily be modified to incorporate filtering methods, such as displaying only the top-k relevant videos.

\subsubsection{Footage Overviewing}
We prompt the LLM to categorize videos into common themes, providing a summary of topics within a user's video collection. The prompt includes a function instruction (\ref{overview_prompt}), followed by the visual narrations of the gallery videos. This prompt is then sent to the LLM to generate the overview, which is subsequently presented in the chat UI for the user to review.

\subsubsection{Idea Brainstorming}
We prompt the LLM to generate creative video editing ideas based on all the user's videos. The prompt structure begins with a function instruction (see \ref{brainstorming_prompt}). If provided, we include creative guidance from the user in the prompt to guide the brainstorming. The creative guidance is extracted as a string argument when LAVE maps action descriptions to function calls (Section \ref{mapping}). If the user does not provide any guidance, it defaults to "general". Following the creative direction, we append the visual narrations of all gallery videos and send the prompt to LLM for completion. Similar to the footage overview, the generated video ideas will be presented in the chat UI.

\subsubsection{Storyboarding}
\label{storyboarding}
LAVE's storyboarding function arranges video clips in a sequence based on a user-provided narrative. Unlike the former functions, it affects only the videos in the timeline. Similar to Idea Brainstorming, the system checks for any creative guidance on the narrative provided by the user, for example, \textit{"Start with my dog's videos then transition to my cat's."} If no guidance is given, the LLM is instructed to create a narrative based on timeline videos. The prompt begins with a function instruction (\ref{storyboard_prompt}), followed by any user narrative guidance, and then the visual narrations of the timeline videos. The output is structured in JSON format, with the key \texttt{"storyboard"} mapping to texts detailing each scene, and \texttt{"video\_ids"} mapping to a list of video IDs indicating the sequence. This format aids downstream processing in parsing the results. Once the execution is complete, the \texttt{"storyboard"} containing scene descriptions will be displayed in the chat UI, and the video order on the timeline will be updated according to \texttt{"video\_ids"}

\subsubsection{Clip Trimming}
LAVE leverages the reasoning and information parsing capabilities of LLMs for trimming video clips. This function analyzes frame captions to identify a video segment that matches a user's trimming command. The function instruction is detailed in \ref{clip_trimming_prompt}. Following the instruction, the user's trimming command and the frame-by-frame captions generated during preprocessing are appended. This compiled prompt is then sent to the LLM for completion. The outputs are also structured in JSON format: \texttt{{"segment": ["start", "end", "rationale"]}}, indicating the start and end frame IDs, as well as the rationale for this prediction. Upon receiving the LLM's response, LAVE updates the UI to display the suggested trim segment and its rationale, thereby aiding the user's understanding of the LLM's decision-making process. Currently, LAVE's trimming precision is one second based on the frame sample rate used in preprocessing. This precision can be adjusted by varying the sampling rates.

\begin{table*}
    \centering
    \caption{Background of the study participants, including their prior experience in video editing, the types of videos they have previously created, and their self-reported understanding of LLM’s capabilities and limitations.}
    \label{participants_info}
    \begin{tabular}{llllll}
        \toprule
        Participants & Editing Experience & Types of Videos Created Before & Understand LLM's Capability\\
        \midrule
        P1 & Proficient & Animated/Explainer & Slightly Disagree \\
        P2 & Proficient & Project & Slightly Agree \\
        P3 & Proficient & Promotional/Action & Slightly Agree \\
        P4 & Beginner & Social Media & Slightly Agree \\
        P5 & Beginner & Project/Presentation & Slightly Agree \\
        P6 & Proficient & Social Media & Slightly Agree \\
        P7 & Beginner & Presentation & Slightly Disagree \\
        P8 & Beginner & (Outdated Experience) & Strongly Agree \\
        \bottomrule
    \end{tabular}
\end{table*}

\subsection{System Implementation}
We implemented the LAVE system as a full-stack web application. The frontend UI was developed using React.js, while the backend server uses Flask. For LLM inferences, we primarily use the latest GPT-4 model from OpenAI. However, for mapping action plans to functions, we employ the \texttt{gpt-4-0613} checkpoint, specifically fine-tuned for function call usage. The maximum context window length for GPT-4 was 8192 tokens during the time we built the system. With these limits, our agent could accommodate and process descriptions from approximately 40 videos in a single LLM call. We use LangChain \cite{langchain}'s wrapper of ChromaDB \cite{chromadb} to construct the vector store. Video pre-processing is performed on a Linux machine equipped with an Nvidia V100 GPU. Finally, we use ffmpeg to synthesize the outcome of users' video edits.

\section{User Study}
We conducted a user study to obtain user feedback on the usage of LAVE. Our study aimed to 1) gauge the effectiveness of LAVE's language augmentation in assisting video editing tasks, and 2) understand user perceptions of an LLM-powered agent within the editing process, particularly its impact on their sense of agency and creativity. For the study, we enlisted participants to use LAVE for editing videos using their own footage, allowing us to test LAVE's functionality and utility across a diverse range of content. In presenting the results, we relate the findings to the design goals of lowering editing barriers with natural language \textbf{(D1)} and maintaining user agency \textbf{(D2)}, highlighting their fulfillment.

\subsection{Participants}
We are interested in understanding how users with diverse video editing experiences receive language-augmented video editing powered by LLM.  To this end, we recruited participants with varying video editing experiences to gather feedback on their perceptions of LAVE. Table \ref{participants_info} presents the background information of each participant.  We recruited eight participants from a tech company, of which three were female, with an average age of 27.6 (STD=3.16). Four of them (P4, P5, P7, P8) identified as beginners in video editing, possessing little to moderate experience. Among the beginners, P8 reported having the least experience, and the last time he edited videos was years ago. Conversely, the other four participants (P1-3, P6) view themselves as proficient, having extensive experience with video editing tools. Among the proficient participants: P1 is a designer but occasionally edits videos for work; P2, having minored in film studies, has been editing videos since high school; P3 runs a YouTube channel and also edits personal family videos; while P6, a PhD student, edits life-log videos for social media weekly. This diverse group allowed us to evaluate LAVE's performance across various editing backgrounds. All participants have had some experience with LLMs. When asked whether they understood the capabilities and limitations of LLMs, participants' responses ranged from "Slightly Disagree" to "Strongly Agree".

\subsection{Study Protocol}
A day before the study, participants were asked to submit a set of videos for pre-processing. They were asked to provide at least 20 clips, each less than a minute long, to fully leverage the system's features. The study duration ranged from 1 to 1.5 hours and was conducted in a quiet environment to minimize distractions. Upon arrival, participants were provided with an overview of the study and a detailed explanation of the LAVE system's features, which took about 15 to 20 minutes. They then engaged with the LAVE system using their own footage, aiming to produce at least one video. Participants had the freedom to explore and produce multiple videos, yet they were required to adhere to a 20 to 30-minute time frame. After their session with the system, participants completed a questionnaire. We solicited feedback on the usefulness and ease of use of each LLM-powered function and the system as a whole. Questions were also posed regarding trust, agency, outcome responsibility, and participants' perceptions of the roles played by the agent. Additionally, we adapted applicable questions from the Creative Support Index \cite{10.1145/2617588}. Finally, users provided their preferences between agent assistance and manual operations for each editing function. All the questions in the questionnaire were based on a 7-point Likert Scale. Subsequently, we conducted semi-structured interviews lasting approximately 20 to 30 minutes. Throughout the study, participants were encouraged to share their thoughts and ask any questions following the think-aloud method \cite{van1994think}. We did not instruct participants to prioritize speed during the study, as it was not the objective. The aim was to observe how users leverage LAVE for video editing and gather feedback.

\begin{figure*}[!t]
    \centering
    \includegraphics[width=1\linewidth]{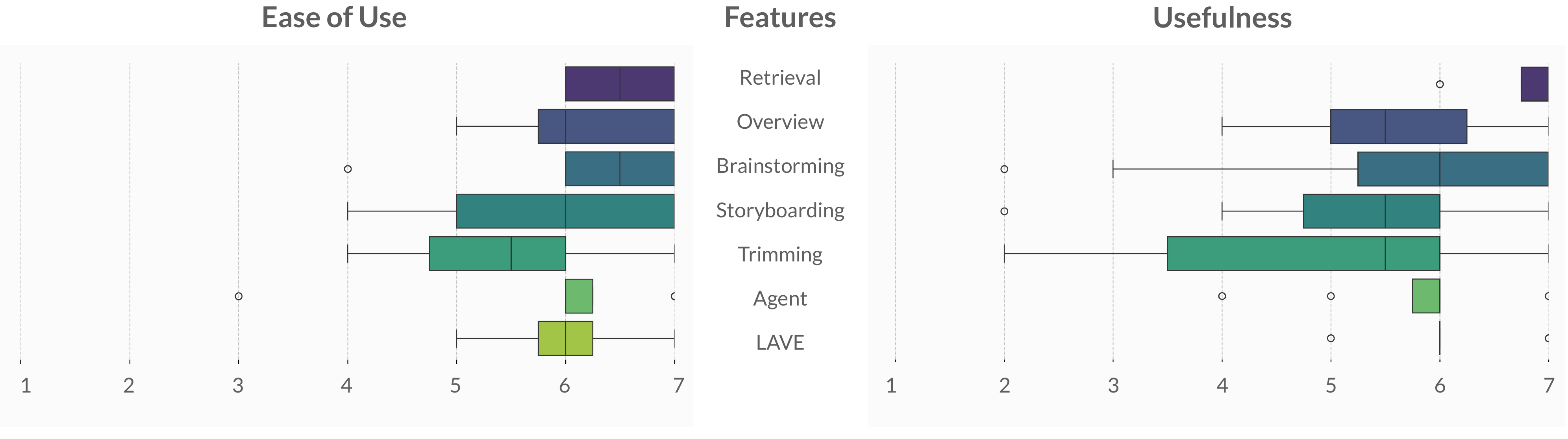}
    \caption{Boxplots showing the ease of use and usefulness of each LLM-powered feature in LAVE, including video retrieval, footage overview, idea brainstorming, storyboarding, and clip trimming. We also solicited feedback on the video editing agent and the overall system. Ratings were based on a 7-point Likert Scale, with 7 indicating "extremely easy to use/useful" and 1 being the opposite. Participants generally found the capabilities of the agent and the full system easy to use. However, variances were observed in usefulness ratings. }
    \label{fig:use-chart}
\end{figure*}

\subsection{Results and Findings}
We summarize the important results and observations obtained from the user study as follows.

\subsubsection{Editing Outcome and General Impressions}
All subjects were able to use LAVE to produce satisfactory video results within the study session with low frustration (Mean=2, STD=1.3). Seven participants rated their satisfaction with the final outcome at 6 out of 7, while participant P2 gave a score of 5. Participants found LAVE enjoyable to use (Mean=6.3, STD=0.5) and expressed a desire to use it regularly (Mean=5.8, STD=0.9). Notably, we observed encouraging results indicating that LAVE reduces editing barriers for inexperienced users \textbf{(D1)}. For instance, P8, who had edited videos only once before, praised the efficiency of using LAVE, stating, \textit{"I really see the value of the tool... in 20 or 30 minutes, you have a really nice video."} This is reinforced by the fact that all beginner users in our study produced satisfactory outcomes in collaboration with LAVE during their first session. The findings underscore LAVE's effectiveness in supporting the video editing process.  

\subsubsection{Constrating LAVE with Existing Editing Tools}
Participants appreciated the novelty of LAVE's editing paradigm. For instance, P3, who is familiar with video editing, commented, \textit{"I think there's nothing like that right now in the market, and I was able to edit a video real quick."} Similarly, P5 made intriguing comments about the role he perceived himself in when using LAVE, saying, \textit{"The system makes me feel like a director, and it's nice to edit videos with a conversational interface because it feels more natural."} He went on to express that he felt he \textit{"operated at a higher level of thinking, which was kind of liberating."} \textbf{(D1)}. We view this as a promising sign for future content editing, where the incorporation of agent-based editing features offers an effective alternative to manual operations.

\subsubsection{Usability and Usefulness}
Participants found the design of LAVE useful and easy to use, as echoed by the overall positive ratings LAVE received, which is illustrated in Figure \ref{fig:use-chart}. However, there were divergent ratings regarding the usefulness of some features. We noticed that negative feedback typically stemmed from two main reasons. Firstly, participants who highly value originality, often proficient editors, tend to prefer maintaining autonomy when conceptualizing video ideas and forming their understanding of videos. As a result, they could be prone to reject conceptualization assistance from the agent. Secondly, due to the stochastic nature of LLMs, outputs of functions such as trimming and storyboarding may not always align with user expectations, leading some participants to rate their usefulness lower. To gain a deeper understanding of the capabilities and limitations of the proposed design, we collected additional user feedback on each LLM-powered editing function, which we discuss below.\\

\noindent \textbf{Video Retrieval:} The feature is unanimously praised for its efficiency in finding relevant videos and received the highest ratings for usefulness. As P1 noted, \textit{"Having the search function is super helpful. Creative people tend to be disorganized, so this can be extremely useful."} Overall, participants were generally surprised by how easily they could find videos using natural language without having to scroll through the corpus. \\

\noindent \textbf{Footage Overview and Video Descriptions:} In soliciting feedback, we combined the footage overview with the pre-generated video narrations (video titles and summaries), as they share the same objective of helping users quickly understand footage content. P6 found the topics and themes outlined in the footage overview useful in aiding him in categorizing the available videos. P2 found the semantic titles helpful and commented on the description's accuracy, stating, \textit{"There was a delight in seeing how it titled everything... Sometimes it was a little wrong, but often it was very right and impressive."} She also highlighted the usefulness of semantic titles over arbitrary IDs commonly assigned by capture devices. \\

\noindent \textbf{Idea Brainstorming:} Brainstorming was found beneficial in aiding the initial conceptualization for the majority of the participants.  As quoted by P3, it can \textit{"spark my creativity."} P8 noted its usefulness in providing starting concepts when he had no ideas. However, not all participants welcomed external input. P2, for example, resisted such suggestions and stated, \textit{"The brainstorming didn't totally work for me. Part of that's because I wouldn't even outsource that to a human assistant."} In addition, P7, having already formed an idea, found brainstorming unnecessary. Moreover, P6, while appreciating the ideas generated by the LLM, expressed concerns about potential bias, commenting, \textit{"It's giving you initial ideas, but it might also bias you into thinking about specific things."}\\

\noindent \textbf{Storyboarding:} The feature was generally viewed as beneficial for sequencing clips. P8 found it supportive, and P4 praised its utility, saying, \textit{"I think it's quite good. I actually had no idea how to sequence them together."} She also valued the narratives provided by the LLM, stating, \textit{"It provided pretty good reasoning for why it sequenced the videos like this."} However, P2 found the reasoning behind some of the storyboards was less grounded,  noting, \textit{"When I asked it for the reason, it would give something grammatically correct but somewhat nonsensical artistically."} This highlights the challenges involved in using LLMs to support storytelling, as they may generate or fabricate implausible narratives. \\

\noindent \textbf{Clip Trimming:} Being able to trim clips based on language commands has generated excitement among users. For instance, while using this feature, P5 remarked, \textit{"It's like telling people what to do, and then they do it; it's kind of amazing."} He was editing a clip that panned from a view inside the car, focusing on the road ahead, to the side window showcasing a group of tall buildings. P5 requested the system to trim five seconds from that transition, and the resulting clip perfectly met his expectations. However, we have observed a non-negligible number of occasions when the LLM inaccurately trimmed videos. This often occurred when users input commands about elements not captured in the system-generated visual narrations, such as brightness or motion. The inaccuracy has led to diminished enthusiasm and lower usefulness ratings, indicating the need for future research.

\subsubsection{Trust, Agency, and Outcome Responsibility}
\begin{figure}[!t]
    \centering
    \includegraphics[width=1\linewidth]{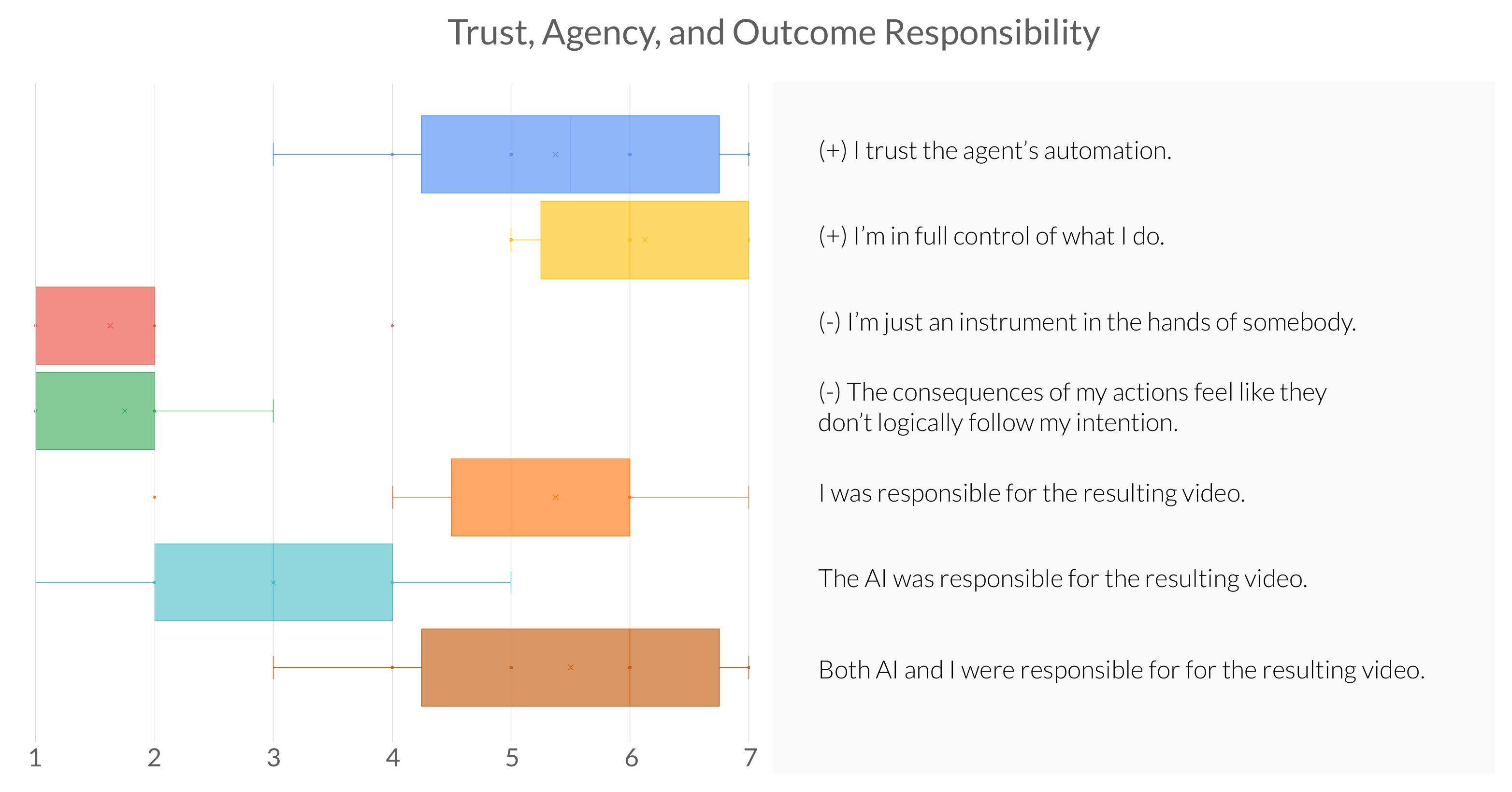}
    \caption{Boxplots showcasing user ratings on trust (first row), agency (second to fourth rows), and outcome responsibility (fifth to seventh rows). All scores use a 7-point Likert scale, where 7 means "strongly agree" and 1 denotes "strongly disagree." Questions marked with (+) indicate that a higher score is preferable., while (-) means the contrary.}
   \label{fig:agency}
\end{figure}

Figure \ref{fig:agency} showcases the user ratings for the questions related to trust, agency, and outcome responsibility. Participants found the automation of LAVE to be generally trustworthy and felt they had a high level of control when using the system, highlighting that they retained agency despite the AI's automation \textbf{(D2)}. When inquiring about responsibility for final outcomes—whether attributed to the AI, the user, or a combined effort—the prevailing sentiment rejected the notion that AI solely influenced the end product. Most agreed they were personally accountable or that it was a joint effort with the AI for the results, except P8, who felt he largely relied on the AI's suggestions.

\subsubsection{Perceptions of the Role of the Editing Agent}
\label{agent_role}
 We further explored how users perceived the role of LAVE's editing agent: whether as an assistant, partner, or leader. Half of the participants regarded the agent as an "assistant" (P2, P3, P7, P8), while the other half perceived it as a "partner" (P1, P4, P5, P6). Notably, none felt as though the AI agent took on a leadership role. Those in the assistant category generally viewed the agent as a responsive tool, following their directives. Conversely, the partner group likened the agent to an equal collaborator, sometimes even equating the experience to engaging with a human peer. P5 remarked, \textit{"When using the tool, I had this partnership with the AI that is kind of similar to having a conversation with somebody, and we're trying to brainstorm ideas."}. In addition, all participants appreciated the ability to have the final say in any editing decision using LAVE, emphasizing their ability to easily refine or dismiss the AI's suggestions \textbf{(D2)}.

\subsubsection{Supporting Creativity and Sense of Co-Creation}
\begin{figure*}[h]
    \centering
    \includegraphics[width=1\linewidth]{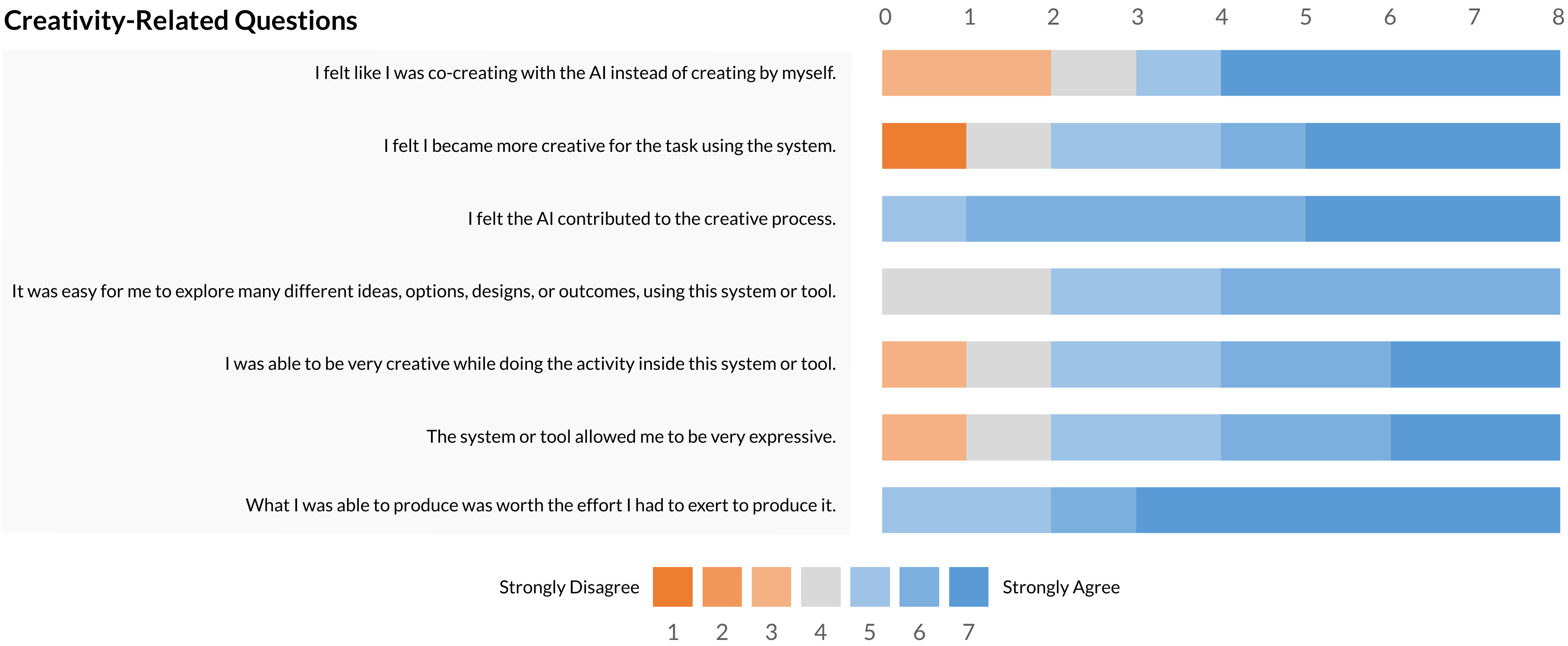}
    \caption{Stacked bar chart of user-reported ratings on questions related to the sense of co-creation and those adopted from the creativity support index \cite{10.1145/2617588}. The horizontal axis in the upper right represents the cumulative number of participants. All responses are rated on a 7-point Likert scale with 1 being "strongly disagree" and 7 being "strongly agree".  Overall, participants expressed positive feelings about their sense of co-creation and the creativity support provided by the LAVE. However, some questions did receive occasional negative feedback, indicating varied perceptions among users. We use a stacked bar graph to highlight the exact proportions of user ratings, particularly those that lean towards disagreement for additional discussion.}
    \label{fig:creativity}
\end{figure*}

  As depicted in Figure \ref{fig:creativity}, users generally were positive about the system's impact on creativity. All users agreed to some extent that AI contributed to the creative process. Furthermore, 6 out of 8 participants believed that the system enhanced their creativity. As P8 mentioned,  \textit{"What's really hindering me from doing video editing is that it's a very creative job, and I feel I lack creativity. This tool addresses that precisely."} \textbf{(D1)}. However, not all participants felt that the system boosted their creativity--P7 was neutral, and P2 strongly disagreed with the statement. When inquiring about users' sense of co-creation, the responses ranged from "slightly disagree" to "strongly agree". Upon analysis, we found that participants who saw the LAVE agent more as a partner (Section \ref{agent_role}) were more likely to feel they were co-creating with AI during the video editing process (Mean=6.5, STD=1). In contrast, those who regarded the LAVE agent merely as an assistant reported a lower sense of co-creation with AI (Mean=4.25, STD=1.9). Lastly, all users were positive that the final results were worth the efforts they exerted in the process, echoing the reported satisfaction with the outcome.
  
\subsubsection{User Preferences for Agent Support}
\label{preference-agent}
We observed a spectrum of assistance that users desired from the LAVE agent. For conceptualization-related tasks, we observed that users who emphasized their creative control showed a tendency to dislike input from the agent (P2, P3). In contrast, P8 expressed a strong desire to intake whatever ideas the agent can offer. For manual operation tasks, a similar trend exists where not all users welcome agent intervention. For example, P8 wants pure automation for storyboarding, while P2 and P7 prefer manually sequencing videos. When it comes to clip trimming, P3, P7, and P8 preferred manual adjustments, emphasizing that the LLM prediction did not fully match their intentions. Overall, the varying preferences across users and tasks indicate future agent-assisted content editing should provide adaptive support rather than a one-size-fits-all approach.

\subsubsection{Users' Mental Models Based on Prior Experience with LLMs}
 We observed that prior experience with LLMs could occasionally influence how users perceived and interacted with LAVE. We found that users with a deeper understanding of LLMs' capabilities and limitations seemed to quickly develop a mental model of how the agent operates. Such users could adapt the way they use the system based on what they believe the LLM can process more effectively. For example, P5 attempted to reuse words from video titles, assuming that the LLM agent would better understand his commands. He also exhibited greater patience when the LLM made errors. Further studying how users develop mental models for LAVE and similar systems, both among those with and without prior experience with LLMs, is an intriguing subject for future research.

\section{Design Implications}
Based on the study findings, we discuss design implications to inform the future design of LLM-assisted content editing systems. \\ 

\noindent \textbf{1. Harnessing Language as a Medium for User Interaction and Content Representation to Enhance Multimedia Editing:}
Our study demonstrates the effectiveness of using natural language as a medium for both user interaction with the system and for representing multimedia content—in our case, representing videos with textual descriptions. The use of language as an interaction medium acts as a liberating factor, reducing manual effort and enhancing user understanding of the editing process. In representing content, language enables us to harness the capabilities of LLMs for versatile processing and editing assistance. A significant implication of our approach extends beyond mere video editing, suggesting that future systems could convert multimedia elements, such as speech, sound events, or even sensory inputs like motion, into textual descriptions. This conversion could allow the systems to leverage the strengths of natural language and LLMs to improve the editing process for a wide range of multimedia content.\\

\noindent \textbf{2. Adapting Agent Assistance to User and Task Variability:} Our research exemplifies how incorporating an LLM agent can improve content editing experiences. However, our study also uncovers that preferences for agent assistance can differ across user groups and the nature of the editing task at hand (Section \ref{preference-agent}). For instance, users who value original ideas may abstain from using brainstorming assistance, while others can be receptive to any suggestions from agents. Moreover, the demand for agent support varies among tasks; particularly, repetitive or tedious tasks are more likely to be delegated to agents. Consequently, we recommend that future systems should provide adaptive agent support, automatically tailored to the preferences of the user and the nature of the task.  These systems could also enable users to activate, deactivate, or customize each assistance as needed. Additionally, we suggest they offer flexibility between agent assistance and manual editing, allowing users to refine AI predictions and correct potential inaccuracies, as demonstrated by LAVE. \\

\noindent \textbf{3. Considering Users' Prior Experience with LLM Agents in System Design:}
Our study suggests that a user's prior experience with LLMs may influence the way they interact with an editing system featuring an LLM agent. Users with a deep understanding of LLMs will likely form a mental model of the agent's functionalities more quickly. Furthermore, those who are adept at using prompting techniques might develop more efficient strategies for interacting with the agent. On the other hand, users who are not well-informed about the capabilities and constraints of LLMs may not utilize the system to its fullest potential. Therefore, it may be beneficial for future systems that incorporate LLM agents to integrate instructional support, such as visual cues that provide feedforward and feedback guidance, especially for users new to LLMs. \\

\noindent \textbf{4. Mitigating Potential Biases in LLM-Assisted Creative Process:}
LAVE's ability to engage users through conversational interactions was perceived by our study participants as both innovative and liberating for video editing, enhancing their ability to operate at a higher level of thinking. However, due to the seemingly human-like nature of LLMs' natural language communication, there is a potential for user mistrust or biases. Some participants highlighted that reliance on LLM suggestions might inadvertently cause them to overlook certain videos they would have considered had they worked independently. Moreover, biases present in LLMs during their training phase \cite{felkner2023winoqueer, venkit2023nationality, yu2023large} have the potential to subtly influence users' creative endeavors. Therefore, it is important to carefully consider the potential for bias introduced by LLMs in the creative process and take steps to mitigate it.

\section{Limitations and Future Work}
LAVE represents an initial step into the emerging field of system design for LLM-assisted content editing. Due to the rapid evolution of LLM research, we acknowledge the transient nature of our current design and implementation. We believe the enduring value of this work lies not in its specific implementation, which may soon evolve, but in its role as the first investigation of the proposed editing paradigm. This sets the stage for the continuous evolution of the field. Below, we discuss limitations that warrant future investigations.

\subsection{Agent Design}
Our agent design draws inspiration from recent work on tool-use agents \cite{karpas2022mrkl, yao2023react, shinn2023reflexion, shen2023hugginggpt}. We anticipate that more sophisticated designs will be proposed to support more robust and versatile interactions. For example, LAVE currently incorporates a single agent with several functions. These functions are executed linearly and do not facilitate back-and-forth discussion within each of them. A potential improvement would be to construct a multi-agent system where each function is represented as a separate agent with which the user can directly interact, for example, a storyboarding agent that engages with users to discuss and clarify desired narratives.  In addition, LAVE's agent requires sequential user approval for actions, a process that future work could vary or make more adaptive. Lastly, while LAVE's agent presently supports only LLM-based editing functions, in practice, it can also incorporate non-LLM-based editing functions, such as those in traditional video editing tools, e.g., visual or sound effects. 

\subsection{Editing Functions}
The editing functions provided by LAVE are not intended to be exhaustive, leaving room for potential improvements. For instance, they could be improved by explicitly distinguishing different aspects of videos, such as objects and activities, and exposing these aspects to the agent for more fine-grained editing control. Future work could also investigate end-user prompting, enabling users to modify or introduce new prompts and tailor their LLM-powered video editing assistance as desired. Lastly, future systems could develop evaluation components to provide an automated feedback loop in the editing process. An evaluation component could be created, for instance, using models capable of assessing visual aesthetics and examining the logic flow in videos. This feature could assess the quality of the editing function's output before it is presented, or it could review the current editing draft to offer users detailed critiques and suggestions.


\subsection{Model Limitations}
\label{llm_limiations}
There are key limitations in using LLMs within LAVE that merit investigation. Firstly, LLMs such as the GPT-4 model \cite{openai2023gpt4}, initially limited to an 8192 token window (now increased to 128k), restrict the amount of video information that can be included in a single prompt. Additionally, LLMs tend to hallucinate, producing grammatically correct yet nonsensical responses, as observed in our user study. Addressing this issue to improve LLM factual accuracy is crucial \cite{tam2022evaluating}. While LLMs cannot currently effectively process video input, recent advancements in VLMs capable of handling image sequences \cite{yang2023dawn} suggest the potential for integrating future VLMs into LAVE. That said, a benefit of our current setup is that when users interact with the system, they may experience quicker processing of textual representation of visuals, as opposed to potentially slower processing of images or videos in real time.

\subsection{User Evaluation}
Our user study evaluated LAVE with eight participants of varying experience, enabling us to understand user perceptions from diverse backgrounds. However, we acknowledge potential limitations in generalizability due to our sample size. Future studies could involve larger participant groups, diverse user backgrounds, or different video editing scenarios to further validate and expand upon our initial findings. Moreover, future work can conduct a quantitative evaluation of the agent's performance and longitudinal studies to examine whether users' behavior with would change as they gain more experience with the proposed editing paradigm.

\section{Conclusion}
We have introduced LAVE, a video editing tool that enables a novel agent-assisted video editing paradigm through LLM-powered assistance and language augmentation. We outlined the system's unique design and implementation, along with its supported functions and language-augmented features. Our user study assessed the effectiveness of LAVE and garnered insights into users' perceptions and reactions to an LLM agent assisting in video editing. Based on the study's findings, we proposed design implications to inform future designs of systems alike. Our work sheds light on the future development of agent-assisted media content editing tools. We are optimistic about the direction and believe we have only begun to scratch the surface of what is possible.

\bibliographystyle{ACM-Reference-Format}
\bibliography{references}

\appendix

\begin{figure*}[!t]
    \centering
    \includegraphics[width=1\linewidth]{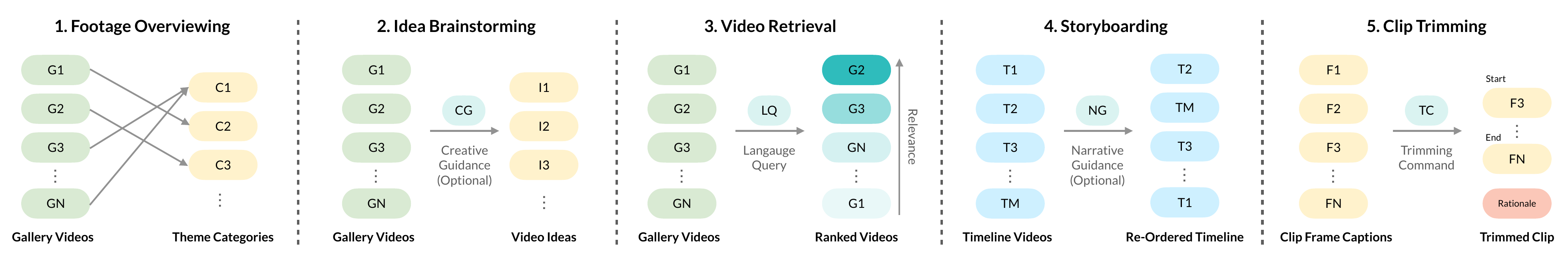}
    \caption{Graph illustrations of the mechanisms of each LLM-based editing function supported by LAVE: \textbf{(1)} In Footage Overview, gallery videos (Gs) are categorized into several common themes or topics (Cs). (2) In Idea Brainstorming, gallery videos (Gs) are used to develop video creation ideas (Is), with the option of creative guidance (CG) being provided by the user. (3) In Video Retrieval, gallery videos (Gs) are ranked based on their relevance to the language query (LQ) extracted from the user's command. Deeper colors in the ranked videos represent higher relevance. (4) In Storyboarding, timeline videos (Ts) are reordered to match narrative guidance (NG) or a storyline optionally provided by the users. If not provided, the model will be asked to generate one itself. (5) In Clip Trimming, captions of each frame in a clip (Fs) will be provided to the model along with the user's trimming command (TC). The function will output the trimmed clip's start and end frame IDs as well as its rationale for the predictions.}
    \label{function-illustrations}
\end{figure*}

\vspace{10em}

\section{Prompt Preambles}
This section contains prompt preambles that instruct the LLM to perform specific editing functions.

\subsection{Footage Overview}
\label{overview_prompt}

\texttt{Summarize the common topics or themes within all the provided videos, or categorize the videos by topic. The overview should be short, informative, and comprehensive, covering all the videos. For each topic or theme, list the titles of the videos that belong to it below.}

\subsection{Idea Brainstorming}
\label{brainstorming_prompt}
\texttt{Use all of the provided videos to brainstorm ideas for video editing. For each idea, specify which video should be used and why. Aim for broad integration of multiple clips; the more comprehensive the integration, the better. Users may provide creative guidance for brainstorming. If the guidance is general, feel free to brainstorm using the videos mentioned below; otherwise, adhere to that guidance.}

\subsection{Storyboarding}
\label{storyboard_prompt}
\texttt{Use all the provided videos to devise a storyboard for video editing. If the user provides creative guidance, follow it closely. Reference videos in the storyboard by their title and ID. The output should be a dictionary with keys: "storyboard" and "video\_ids". The "storyboard" key maps to a string detailing each scene in the storyboard, in the format of "Scene X: <Video Title> (ID=X), <rationale for scene placement>". The "video\_ids" key maps to a sequence of video IDs as referenced in the storyboard. Ensure all input videos are included in the output.}

\subsection{Clip Trimming}
\label{clip_trimming_prompt}
\texttt{Given video frame captions with timestamps, where each description represents 1 second of video, and the user's trimming command, determine the new start and end timestamps for the trimmed clip. If a specific clip length constraint is mentioned, adhere to it. The expected output is a Python dictionary formatted as: Final Answer: \{"segment": ["start", "end", "rationale"]\}. Both "start" and "end" should be integers. If no segment matches the user's command, "segment" should contain an empty list. Prioritize longer segments when multiple qualify.} 

\end{document}